\documentclass[11pt]{elsart}

\usepackage[dvips]{graphicx}
\usepackage{array}
\usepackage{amssymb}
\usepackage{amsmath}
\usepackage{hhline}
\usepackage{longtable}
\usepackage{dcolumn}
\usepackage{bm}
\usepackage{subfigure}

\begin{document}
\begin{frontmatter}

\title{Local-world evolving networks with tunable clustering}
\author[ad1,ad2]{Zhongzhi Zhang\corauthref{zz}}
\corauth[zz]{Corresponding author.} 
\ead{zhangzz@fudan.edu.cn}
\author[ad3]{Lili Rong}
\ead{llrong@dlut.edu.cn}
\author[ad4]{Bing Wang} 
\author[ad1,ad2]{Shuigeng Zhou}
\ead{sgzhou@fudan.edu.cn}
\author[ad5]{Jihong Guan}
\ead{jhguan@mail.tongji.edu.cn}
\address[ad1]{Department of Computer Science and Engineering, Fudan
University,\\Shanghai 200433, China}%
\address[ad2]{Shanghai Key Lab of Intelligent Information Processing, Fudan
University,\\ Shanghai 200433, China}%
\address[ad3]{Institute of Systems Engineering, Dalian University of Technology,
\\ Dalian 116024, Liaoning, China}%
\address[ad4]{Department of Applied Mathematics , Dalian University of Technology,
\\ Dalian 116024, Liaoning, China}%
\address[ad5]{Department of Computer Science and Technology, Tongji University,
\\4800 Cao'an Road, Shanghai 201804, China}%

\begin{abstract}
We propose an extended local-world evolving network model including
a triad formation step. In the process of network evolution, random
fluctuation in the number of new edges is involved. We derive
analytical expressions for degree distribution, clustering
coefficient and average path length. Our model can unify the generic
properties of real-life networks: scale-free degree distribution,
high clustering and small inter-node separation. Moreover, in our
model, the clustering coefficient is tunable simply by changing the
expected number of triad formation steps after a single local
preferential attachment step.


\end{abstract}

\date{}
\end{frontmatter}

\section{Introduction}
Complex networks~\cite{AlBa02,DoMe02,Ne03,BoLaMoChHw06,BoSaVe07}
describe a number of systems in nature and society, such as Internet
\cite{FaFaFa99}, World Wide Web \cite{AlJeBa99}, metabolic networks
\cite{JeToAlOlBa00}, protein networks in the cell \cite{JeMaBaOl01},
co-author networks \cite{Ne01a} and sexual networks
\cite{LiEdAmStAb01}. Many real-life networks share three apparent
features: (a) The degree distribution of nodes is scale-free, i.e.,
it follows a power law. (b) The clustering coefficient is high. Two
nodes having a common neighbor are far more likely to be linked to
each other than are two nodes selected randomly. (c) The average
path length (APL) is small. That is, the expected number of edges
needed to pass from one arbitrarily selected node to another is low.

How to model real complex networks with these three properties?
Traditionally the study of complex networks has been the scope of
graph theory. While graph theory initially focused on regular
graphs, since the 1950s large-scale networks with no apparent design
principles have been depicted as random graphs \cite{ErRe59,ErRe60},
proposed as the simplest and most straightforward realization of a
complex network. In the past ten years, 
scientists have found that most real-life networks are neither
completely regular nor completely random, but have three properties
above. So they proposed some new models to depict real-life
networks. Among them, two are the most well known. One is the
small-world network model \cite{WaSt98} proposed by Watts and
Strogatz (WS) in the year of 1998, which interpolates between
regular and random graphs and has two properties of high clustering
and short APL. The other is scale-free network model
\cite{BaAl99,BaAlJe99} with power-law degree distribution and low
APL presented by Barab\'asi and Albert (BA).

Although the two pioneering (WS and BA) models played an important
role in network science and started an avalanche of research on
complex networks \cite{NeWa99,BaAm99,BaWe00,DoMeSa00,KrReLe00},
neither of them can completely describe the three common
characteristics of real-life networks. After that, a great number of
attempts have been made at constructing
models~\cite{HoKi02,KlEg02a,KlEg02b,SaKa04,AnHeAnSi05,DoMa05,ZhYaWa05,ZhCoFeRo05,ZhRoCo05,ZhRoZh06}
with the three properties coinciding with real-life networks. Holme
and Kim extended the BA model to include a triad formation step
\cite{HoKi02}. Klemm and Egu\'iluz introduced a growing network
model based on a finite memory of nodes \cite{HoKi02,KlEg02a}.
Saram\"aki and Kaski presented an undirected scale-free network
model generated by random walkers~\cite{SaKa04}. Andrade \emph{et
al. }\cite{AnHeAnSi05} introduced Apollonian networks on the basis
of the problem of Apollonian packing, which were also proposed by
Doye and Massen \cite{DoMa05}. Apollonian networks have received
much attention from the scientific community. Zhou\emph{ et al.}
presented a simple rule generating random two-dimensional Apollonian
networks \cite{ZhYaWa05}. Zhang\emph{ et al.} offered a simple
general algorithm producing high-dimensional deterministic and
random Apollonian networks~\cite{ZhCoFeRo05,ZhRoCo05,ZhRoZh06}.

All above models may capture some mechanisms responsible for the
three common traits shared by real-life networks, but they have
ignored some other significant factors. For example, in various
real-life networks such as World Trade Web \cite{SeBo03,LiYCh03}
when a new node enters the system, it doesn't have the globe
information of all existing nodes, so preferential attachment
mechanism only works on the local-world of the new node. To better
understand and describe this real-life phenomenon, Li and Chen (LC)
proposed a local-world evolving network model~\cite{LiCh03}, which
has found applications in many fields such as
Internet~\cite{ChFaLi05} and society~\cite{WaTaZhXi05}. However, LC
model has a low clustering coefficient.

In this paper, in order to portray real-life network more
appropriately, we present a local-world evolving network model with
changeable local-world size and tunable clustering, which can
capture both the mechanism of local preferential
attachment~\cite{LiCh03} and triad formation~\cite{HoKi02}. In the
model there is a random fluctuation in the number of new edges
acquired by the network which is more realistic
\cite{LilaYeDa02,LilaYe02}. We analyze the geometric characteristics
of the model both analytically and numerically. The analytical
expressions are in good agreement with the numerical simulations.
Our model has the three common features of real-life networks.
Moreover, it represents a transition between exponential scaling and
power-law, so it may depict some real-networks such as scientific
collaboration network \cite{Ne01b} whose degree distribution is
neither power-law nor exponential.


\section{The LC Local-world Evolving Network}

Two ingredients, i.e. growth and preferential attachment in
local-world, inspired Li and Chen to introduce the LC model
\cite{LiCh03} for dynamical evolving networks. The LC model can
capture the localization of real-life networks, and its generation
algorithm is as follows:

(1) Initial condition: The network has a small number $(m_{0})$ of
nodes and small number $(e_{0})$ of edges. And then we perform the
following two steps:

(2) Growth: At every time step, we add one node $v$ with $m$
($m<m_{0})$ edges to the existing network.

(3) Determining local-world: Randomly choose $M$ nodes from the
existing network, which are considered as the ``Local-world" of the
new node $v$.

(4) Local Preferential attachment (LPA): The node $v$ connects to
$m$ different nodes in its local-world determined in step (3). We
assume that the probability $\Pi _{Local}(k_{i})$  that node $v$
will be connected to an old node $i$, which is in the local-world of
node $v$, depends on the degree $k_{i}$ of node $i$:
\begin{equation}
\Pi _{Local}(k_{i})=\frac{M}{m_{0}+t} \frac{k_{i}}{\sum _{Local}
k_{j}}.
\end{equation}

In the case of $m<M<m_{0}+t$, the LC model represents a transition
between power-law and exponential scaling networks. In particular,
the original BA model \cite{BaAl99} is a special case of this
local-world evolving network model.

\section{Extended local-world evolving networks with changeable
local-world size and tunable clustering}

\begin{figure}
\begin{center}
\includegraphics[width=11cm]{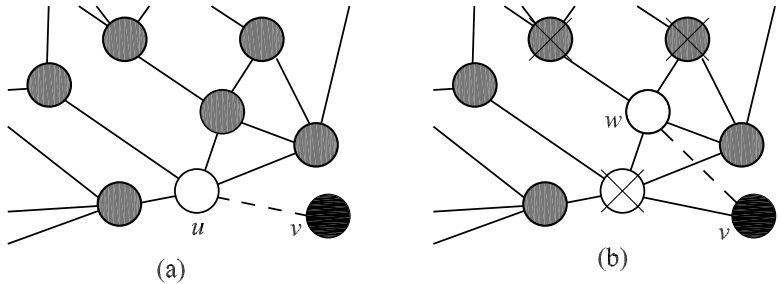}
\caption{Sketch of local preferential attachment and triad
formation. In the LPA step (a) with a probability proportional to
its degree a node $u$ is selected to link to the new node $v$. In
the TF step (b), we choose randomly one neighborhood $w$ of node $u$
that was selected connecting to the new node $v$ in the previous LPA
step, then we add a link between $w$ and $v$. $\times$ symbolizes
``not allowed to attach to". } \label{fig:cluster}
\end{center}
\end{figure}

The LC model captures a common characteristic of many real-life
networks: nodes have local-world connectivity. However, the
clustering coefficient of the LC model approaches zero when the
network size is large, which will be addressed in the following
section. To incorporate the high clustering, we make use of the
method introduced by Holme and Kim~\cite{HoKi02} to modify the LC
model by adding an additional triad formation (TF) step: In the
previous LPA step if there is an edge connecting the new node $v$
and one existing node $w$, then we add one more edge from $v$ to a
randomly selected neighbor of $w$ with a given probability (see Fig.
\ref{fig:cluster}). If all neighbors of $w$ have been connected to
$v$, do an LPA step instead.

 The detailed description is as follows: at
every time step, when a new node $v$ enters the existing network, we
perform an LPA step first, and then with the probability $p$ we
implement a TF step. In succession, we carry out an LPA step
followed by a TF step with the probability $p$. After this process
repeats $m$ times we go to the next time step. It is worth noticing
that in our model an LPA step is always followed by a TF step with
probability $p$, which we take as the control parameter in our
model. So in the present model, at every time step there are $m$ LPA
steps and $p$-dependent TF steps between 0 and $m$ with expectation
$mp$. That is to say, when new nodes are added to the network at
different time steps, the number of new edges is generally not
constant \cite{LilaYe02}. After $t$ time steps, the model develops
to a network with $N_{t}=m_{0}+t$ nodes and expected
$E_{t}=(1+p)mt+e_{0}$ edges. Then the average node degree is
$\langle k\rangle_t=2\,E_{t}/N_{t}$ which is approximately equal to
$2m(1+p)$ for infinite network size.

Additionally, with the network growth more information is available
for the new nodes, so the size $M$ of local-world increases with
time. Thus, different from the LC model, we allow for a change in
$M$, which is denoted here $M_{t}$. We assume that
$M_{t}=a(m_{0}+t)+b$ and limit $m \leq M_{t}\leq m_{0}+t$.

Note that many real-life networks exhibit such an evolving mechanism
as described in our model. For example, in the network of scientific
citations, a new manuscript is more likely to cite well-known and
thus much-cited publications than less-cited and consequently
less-known papers in the same field of the manuscript (i.e. its
local-world). Moreover, with the lapse of time, there are more
papers available in the field for new manuscripts to refer, so the
size of local-world increases with time. On the other hand, in the
content of citations a not untypical scenario is that after citing a
few famous references an author may simply cite secondary references
from the famous ones (TP steps).

It is evident that at every time step, the parameters in the our
model always meet the following conditions: $m \leq M_{t}\leq
m_{0}+t$ and $0 \leq p\leq 1$. So there are at least three limiting
cases as below:

Case A: When $p=0$, and $M_{t}=m$, it is a growing network with
uniform attachment which is the same case of model A in
Ref.~\cite{BaAlJe99}.

Case B: When $p=0$, and $M_{t}=t+m_{0}$, the local-world of the new
node is the whole network, our model reduces to the original BA
model \cite{BaAl99,BaAlJe99}.

Case C: When $p=0$, $M_{t} =const$ and $m \leq M_{t}\leq m_{0}+t$ ,
the model is reduced to the LC model \cite{LiCh03}.

As discussed in Section 2, when $m \leq M_{t}\leq m_{0}+t$, the LC
model represents a transition between power-law and exponential
scaling networks. It is the same with our model, because the LC
model is a special case of ours. So our considered model may depict
some real-life networks whose degree distribution is neither
power-law nor exponential.

\section{Analytical calculation of relevant network parameters}
Topology properties are of fundamental significance to understand
the complex dynamics of real-life systems. Here we focus on three
important characteristics: degree distribution, clustering
coefficient, and average path length.
\subsection{Degree distribution}
Degree distribution is one of the most important statistical
characteristics of a network. Below we will show that the size of
local-world has a significant effect on the degree distribution.
\subsubsection{Case of $M_{t}\gg m$}
If the local-world scale has $M_{t}\gg m$, our model has a power-law
degree distribution, similar to the BA network
\cite{BaAl99,BaAlJe99}. We can interpret this by calculating
analytically based on the mean-field approach in Refs.
\cite{BaAlJe99,HoKi02,LiCh03}. We assume that the degree $k_{i}$ of
node $i$ is continuous, and thus the probability given by Eq. (1)
can be interpreted as a continuous rate of change of $k_{i}$. In an
LPA step, node $i$ increases its degree with the rate
\begin{equation}\label{kiLPA}
\frac{\partial k_{i}}{\partial
t}=\frac{M_{t}}{m_{0}+t}\frac{k_{i}}{\sum_{Local} k_{j}}.
\end{equation}
 For a TF step, we can gain the average increase of $k_{i}$ via the
probability given by the following equation
\begin{equation}\label{kiTP}
\frac{\partial k_{i}}{\partial t}=\frac{M_{t}}{m_{0}+t}\sum_{n\in
\Omega} \frac{k_{n}}{\sum_{local}
k_{j}}\frac{1}{k_{n}}=\frac{M_{t}}{m_{0}+t}\frac{k_{i}}{\sum_{Local}
k_{j}},
\end{equation}
where $\Omega$ is the set of neighbors of node $i$, and $k_{i}$ is
the number of nodes in $\Omega$.

Similar to the fluctuation in Refs. \cite{LilaYeDa02,LilaYe02}, here
the fluctuation of triad formation steps also has little impact on
the growth dynamics $k_{i}(t)$ of node $i$ and degree distribution
$P(k)$ when the network size is large enough. So we can suppose that
in one time step we perform $m$ LPA steps and $mp$ TF steps on
average. From Eqs.~(\ref{kiLPA}) and ~(\ref{kiTP}) the total rate
per time step is expressed as
\begin{equation}\label{ki01}
\frac{\partial k_{i}}{\partial
t}=m\frac{M_{t}}{m_{0}+t}\frac{k_{i}}{\sum_{Local}
k_{j}}+mp\frac{M_{t}}{m_{0}+t}\frac{k_{i}}{\sum_{Local}
k_{j}}=\frac{m(1+p)M_{t}}{m_{0}+t}\frac{k_{i}}{\sum_{Local} k_{j}}.
\end{equation}
We assume that the cumulative node degree in the
local-world~\cite{LiCh03} meets the following expression
\begin{equation}\label{localdegree}
\sum_{Local} k_{j}=\langle k \rangle M_{t},
\end{equation}
where $\langle k \rangle$ is the average degree of all the nodes in
the network. Substituting Eq.~(\ref{localdegree}) into
Eq.~(\ref{ki01}), we obtain
\begin{equation}\label{ki02}
\frac{\partial k_{i}}{\partial t}=\frac{m(1+p)M_{t}}{m_{0}+t}
\frac{k_{i}}{\frac{2[(1+p)mt+e_{0}]M_{t}}{m_{0}+t}}\approx\frac{k_{i}}{2t},
\end{equation}
which has the same form as the original BA model
\cite{BaAl99,BaAlJe99}. The solution of this equation, with the
initial condition that node $i$ was added to the system at time
$t_{i}$ with the expected value of connectivity $k_{i}(t_{i})=
m(1+p)$, is
\begin{equation}\label{ki03}
k_{i}(t)=m(1+p)\left(\frac{t}{t_{i}}\right)^{0.5},
\end{equation}
which results in the power-law degree distribution of form
\cite{DoMeSa00,KrReLe00}
\begin{equation}\label{Pk01}
P(k)=\frac{2m(1+p)[m(1+p)+1]}{k(k+1)(k+2)},
\end{equation}
which at the limit $m(1+p)\gg1$ and $k\gg1$ can be written in the
common form $P(k)=2(1+p)^{2}m^{2}k^{-3}$ \cite{SaKa04}.

\begin{figure}
\centering
\begin{tabular}{cc}
\subfigure [$p=0.0$]
{
    \includegraphics[width=6cm]{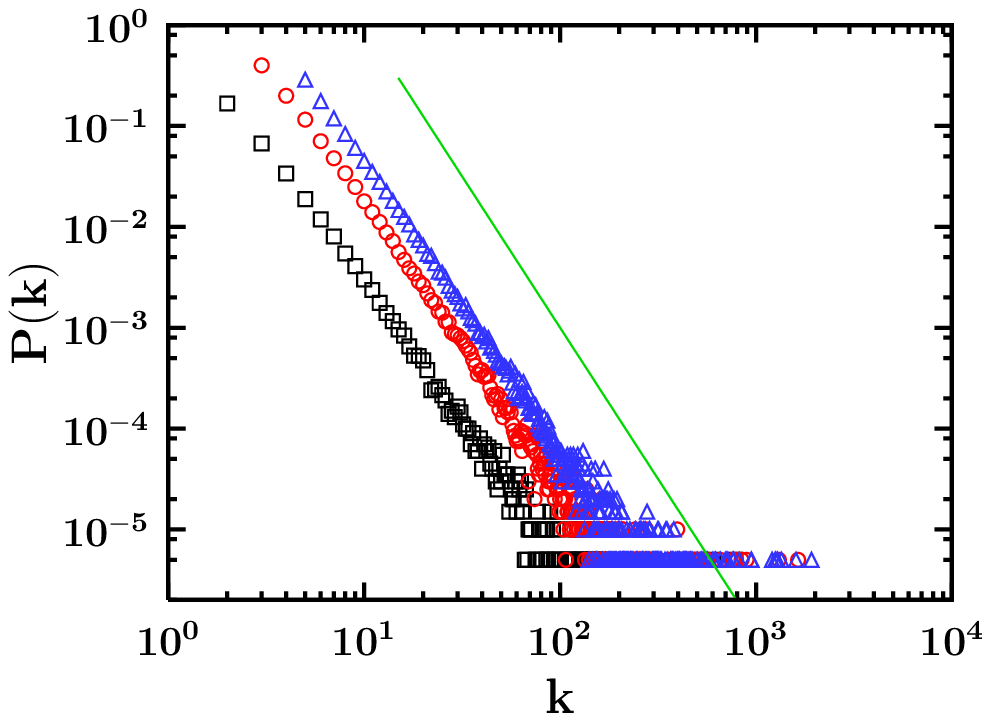}
}
\subfigure[$p=0.3$] 
{
    \includegraphics[width=6cm]{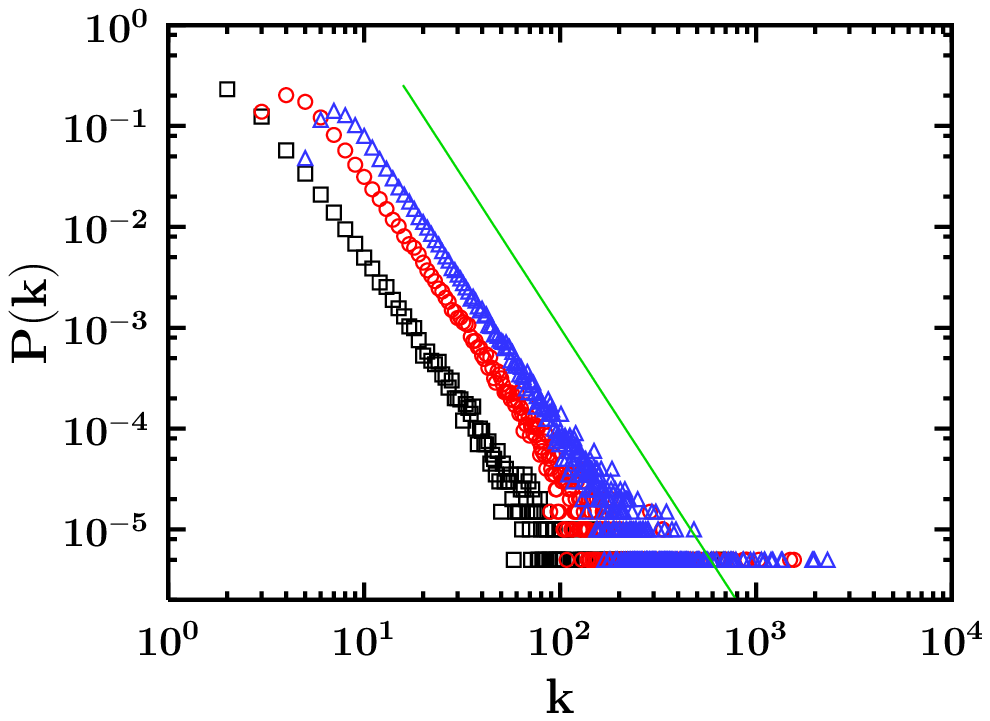}
}\\
\subfigure[$p=0.6$] 
{
    \includegraphics[width=6cm]{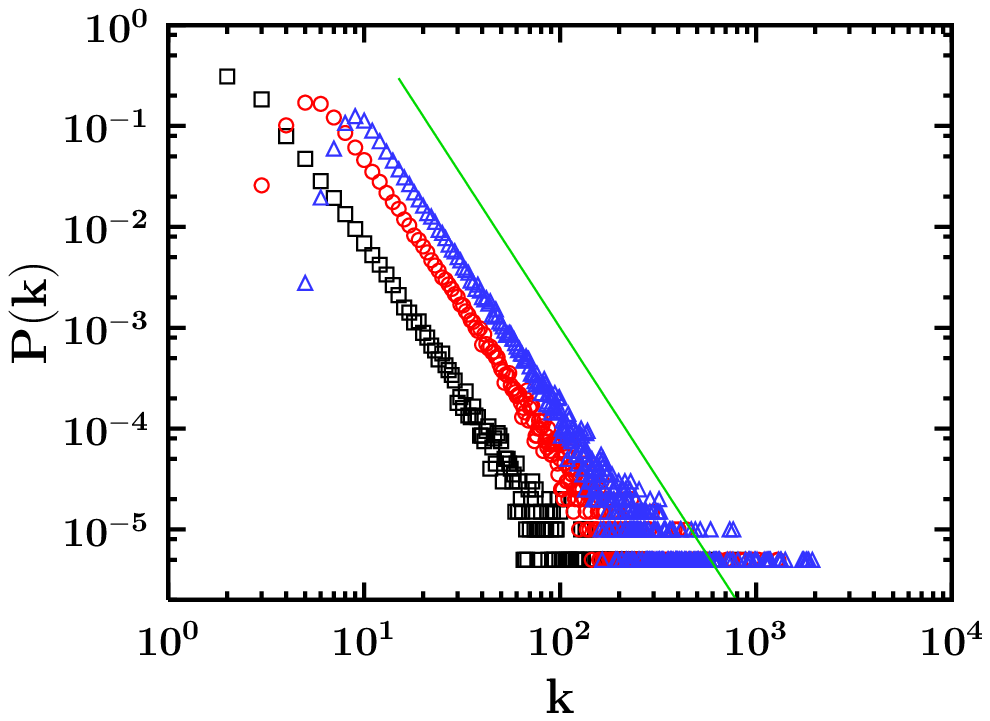}
}
\subfigure[$p=1$] 
{
    \includegraphics[width=6cm]{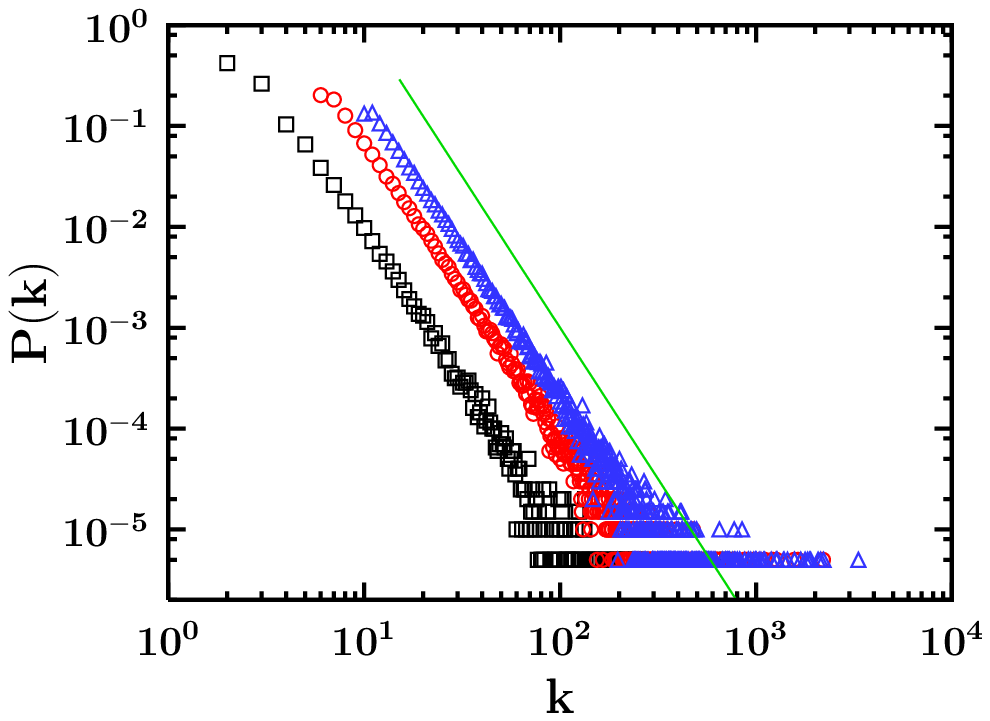}
}
\end{tabular}
\caption{Degree distribution $P(k)$ versus $k$ on a logarithmic
scale. The size of networks is 200,000. The open squares, circles
and triangles denote the cases of $m$=1, 3, 5, respectively. The
slope of all the four straight lines is -3. }
\end{figure} \label{fig:degree01}

In Fig. 2, we report the degree distribution at various values of
$p$ and $m$. In the process of simulation, the local-world size
$M_{t}$ scales as $M_{t}=0.3(m_{0}+t)+m$. From Fig.~2, one can
easily see that the simulation results agree very well with the
theoretical ones. Comparing (b) and (c) with (a) and (d), we observe
that for small values of $k$, there is a deviation from power-law
behavior in (b) and (c), which originates from the fluctuation in
the number of new links acquired by the system (see Refs.
\cite{LilaYeDa02,LilaYe02}). It should be noted that many real-life
networks such as the World Wide Web \cite{AlJeBa99}, the actor
collaboration graphs~\cite{WaSt98} and scientific collaboration
networks~\cite{Ne01b} indeed exhibit this phenomenon of deviation
from power-law to some degree for small $k$ values.

\subsubsection{Case of $M_{t} = m$}
Now, we investigate the case of $0 <p \leq 1$ and $M_{t}= m$.
Obviously, for this particular case, the LPA is reduced to uniform
attachment~\cite{BaAlJe99} which means that the new node is
connected to existing nodes in the network with equal probability,
and a TP step corresponds to a random walk in the network which
implies that an old node acquires a new link with a probability
proportional to its degree~\cite{SaKa04}. Thus, when $t\gg m_{0}$,
the total change rate of degree $k_{i}$  of an old node becomes
\begin{equation}\label{eq:kic2}
\frac{\partial k_{i}}{\partial
t}=m\frac{1}{m_{0}+t}+mp\frac{k_{i}}{2[m(1+p)t+e_{0}]}\approx m\,
\frac{1}{t}+\frac{p}{1+p}\,\frac{k_{i}}{2t}.
\end{equation}
The solution of Eq.~(\ref{eq:kic2}), with initial condition
$k_{i}(t_{i})=m\,(1+p)$, has the form
\begin{equation}\label{eq:ki2}
k_{i}= \frac{m(p+2)(p+1)}{p} \left ( \frac{t}{t_{i}} \right
)^{\frac{p}{2(1+p)}}-\frac{2m(1+p)}{p}.
\end{equation}
Using Eq.~(\ref{eq:ki2}), the probability that a node has a degree
$k_{i}(t)$ smaller than $k$, $P(k_{i}(t)<k)$, can be written as
\begin{eqnarray}\label{pki}
P(k_{i}(t)<k)&=P  \left[ t_{i} > \left (
\frac{p(1+p)m+2m(1+p)}{pk+2m(1+p)} \right )^{\frac{2(1+p)}{p}}\,t
\right ]\nonumber \\
&= 1-P  \left[ t_{i} \leq \left ( \frac{p(1+p)m+2m(1+p)}{pk+2m(1+p)}
\right )^{\frac{2(1+p)}{p}}\,t \right ].
\end{eqnarray}
Assuming that we add the nodes at equal time intervals to the
system, the probability density of $t_{i}$ is~\cite{BaAlJe99}
\begin{equation}
P_{i}(t_{i})=\frac{1}{m_{0}+t}\approx \frac{1}{t}.
\end{equation}
Thus, Eq.~(\ref{pki}) may be rewritten as
\begin{equation}
P(k_{i}(t)<k)=1-\left ( \frac{p(1+p)m+2m(1+p)}{pk+2m(1+p)} \right
)^{\frac{2(1+p)}{p}}.
\end{equation}
Then the degree distribution $P(k)$ can be obtained using
\begin{eqnarray}\label{Pk2}
P(k)&=& \frac{\partial P(k_{i}(t)<k)}{\partial k}\nonumber \\
&=&\frac{2(1+p)}{p^{2+2/p}}\,\left [ m(1+p)(2+p) \right
]^{\frac{2(1+p)}{p}} \left [k+\frac{2m(1+p)}{p} \right ]
^{-(3+\frac{2}{p})}.
\end{eqnarray}
Equation~(\ref{Pk2}) exhibits the extended power-law form as
\begin{equation}\label{Pk3}
P(k)\sim (k+ \kappa) ^{-\gamma},
\end{equation}
where $\kappa =\frac{2m(1+p)}{p}$ and $\gamma=3+\frac{2}{p}$, which
depends on $p$ and is larger than 5. When $k$ is much larger than
$\kappa$, Eq.~(\ref{Pk3}) is reduced to the scale-free form
$P(k)\sim k^{-\gamma}$. Conversely, when $k$ is much smaller than
$\kappa$, we have
\begin{equation}
\ln[P(k)]\sim -\gamma \ln (k+ \kappa)= -\gamma \left [ \ln \left (1+
\frac{k}{\kappa} \right ) +\ln \kappa \right ] \sim -\gamma \left
[\frac{k}{\kappa} +\ln \kappa \right ].
\end{equation}
Thus, we can obtain
\begin{equation}\label{Pk4}
P(k)\sim \frac{1}{\kappa ^{\gamma}}\exp \left ( -\frac{\gamma
k}{\kappa}\right ),
\end{equation}
which shows that Eq.~(\ref{Pk3}) reduces to the exponential form
$P(k)\sim \exp \left ( -\frac{\gamma k}{\kappa}\right )$.

\begin{figure}
\centering
\begin{tabular}{cc}
\subfigure [$p=0.2$]
{
    \includegraphics[width=6cm]{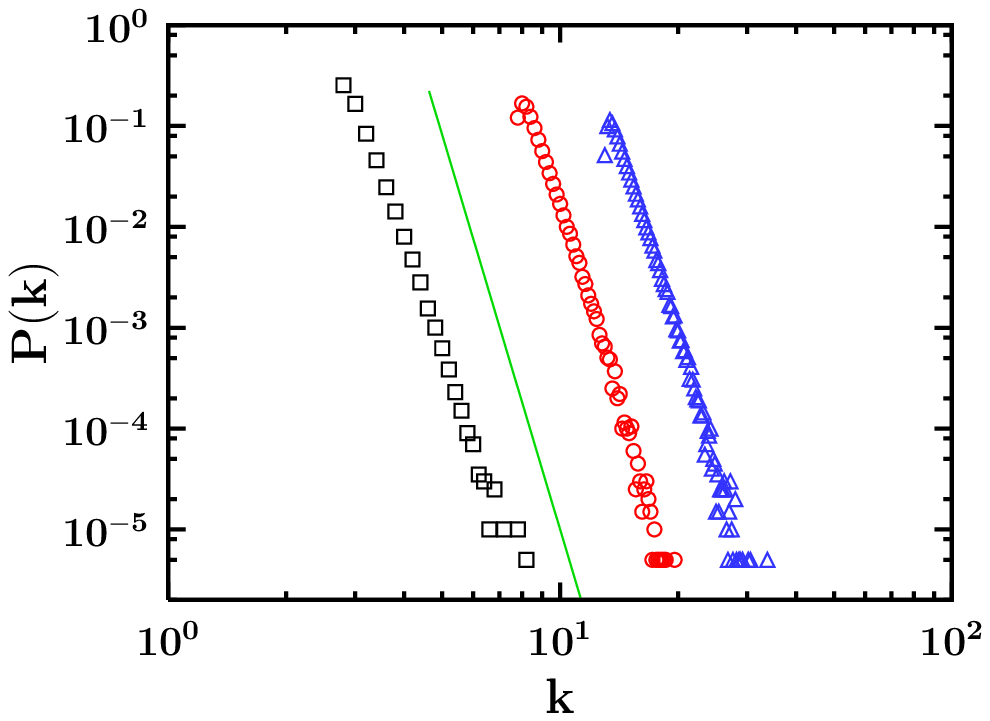}
}
\subfigure[$p=0.4$] 
{
    \includegraphics[width=6cm]{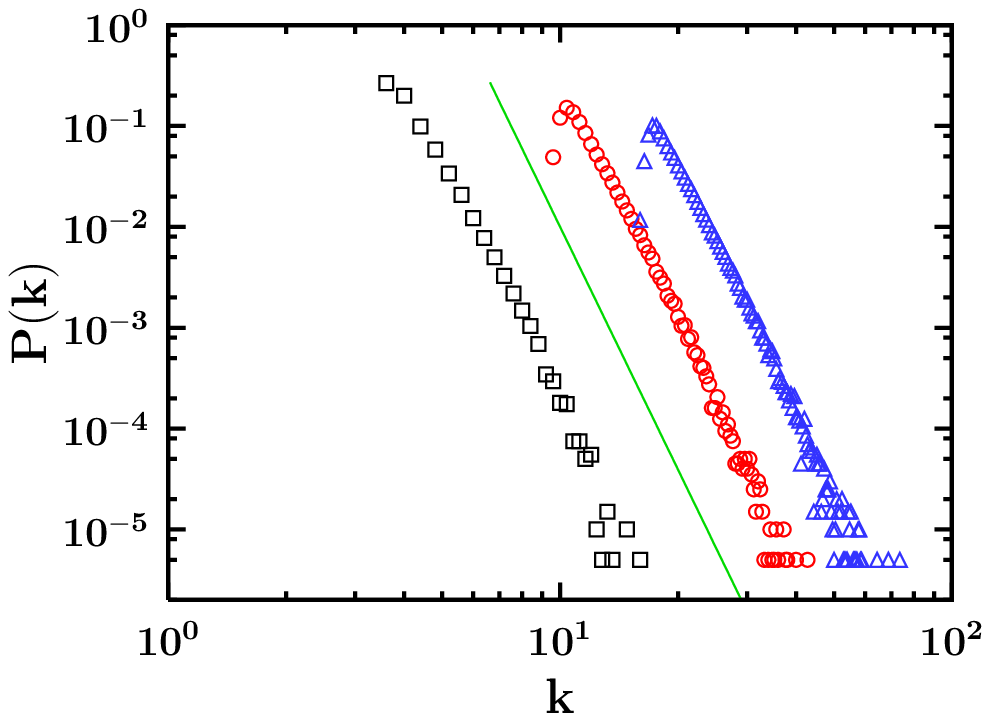}
}\\
\subfigure[$p=0.6$] 
{
    \includegraphics[width=6cm]{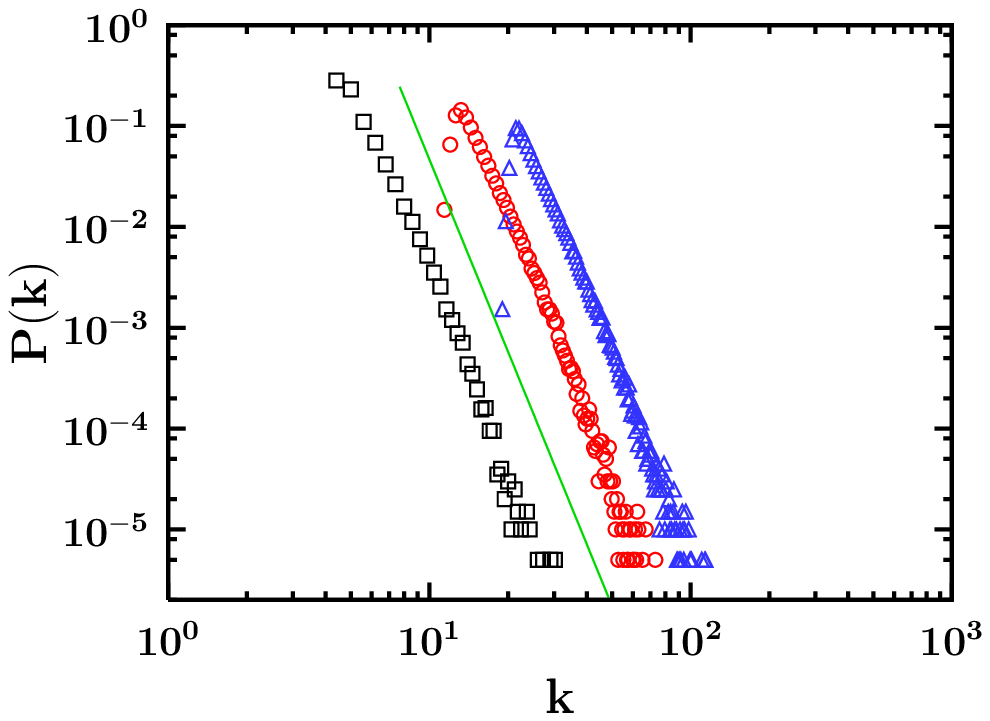}
}
\subfigure[$p=0.8$] 
{
    \includegraphics[width=6cm]{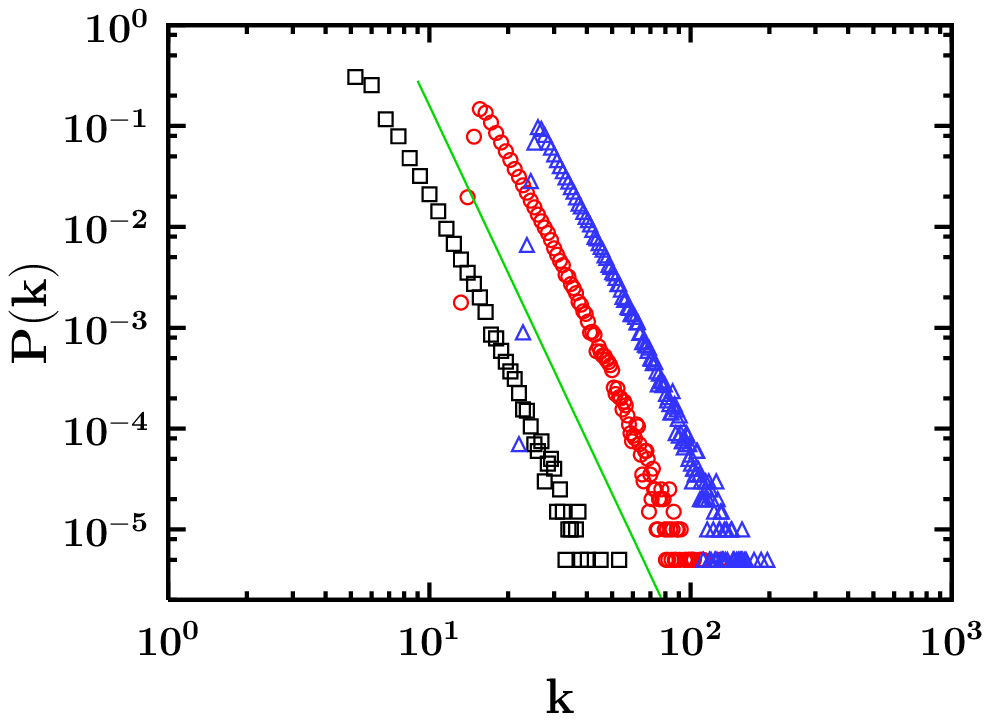}
}
\end{tabular}
\caption{Degree distribution $P(k)$ versus $k$ on a logarithmic
scale . The size of networks is 200,000. The open squares, circles
and triangles denote the cases of $m$=1, 3, 5, respectively. The
four straight lines are the theoretic results predicted by Eq.~(\ref{Pk3}). }
\end{figure}\label{fig:degree02}

From above discussion, we can easily see that the network in this
limiting case follows a stretched exponential distribution, which
has been observed in many real-life
systems~\cite{LaSo98,ZhChetal06}, such as public transport networks
and actor collaboration networks. It should be noted that, for
$M_{t}\approx m$, the network has obviously similar degree
distribution as that of case $M_{t} = m$. In Fig. 3, we show the
degree distribution for $M_{t} = m$ and various values of $p$, the
simulations are consistent with our theoretical prediction.

\subsection{Clustering coefficient}
Most real-life networks show a cluster structure which can be
quantified by the clustering
coefficient~\cite{AlBa02,DoMe02,Ne03,BoLaMoChHw06,BoSaVe07}. The
clustering coefficient of a node gives the relation of connections
of the neighborhood nodes connected to it. By definition, clustering
coefficient \cite{WaSt98} $C_{i}$ of a node $i$ is the ratio of the
total number $e_{i}$ of existing edges between all its $k_{i}$
nearest neighbors and the number $k_{i}(k_{i}-1)/2$ of all possible
edges between them, i.e. $C_{i}=2e_{i}/[k_{i}(k_{i}-1)]$. The
clustering coefficient of the whole network is the average of all
individual $C_{i}'s$.

Using the mean-field rate-equation theory \cite{SzZlKe03} we can
calculate $C_{i}$ analytically. Here we place our emphasis on the
case of $M_{t}\gg m$. Figure 4 illustrates the main microscopic
mechanisms increasing $e_{i}$: (a) node $i$ is connected to the new
node in an LPA step, which is potentially followed by one TF step;
(b) in an LPA step the new node attaches to one of the neighbors of
$i$, and then in one of the subsequent TF steps the new node
conversely gets linked to $i$; (c) in an LPA step node $i$ is
connected to the new node and in another LPA step a neighbor of $i$
is also selected for connection to the new node; (d) in a TF step
node $i$ is connected to the new node and in another TF step a
neighbor of $i$ is also selected for connection to the new node. (e)
node $i$ is connected to the new node in an LPA step, and in the
potential TF steps which follow LPA steps when the new node connects
to the neighbor nodes of $i$, the new node gets linked to $i$. Here
we exclude secondary triangle formation that takes place if two TF
steps from the new node form a triangle composed of two of $i's$
neighbors and the new node, which has little effect on the
clustering of node $i$. So the rate equation for $e_{i}$ reads
\begin{eqnarray}\label{ei01}
\frac{\partial e_{i}}{\partial t}=m\frac{k_{i}}{2m(1+p)t}p&+&
m\sum_{n\in
\Omega}\frac{k_{n}}{2m(1+p)t}\frac{1}{k_{n}}p \nonumber \\
&+& m\frac{k_{i}}{2m(1+p)t}(m-1)\sum_{n\in
\Omega}\frac{k_{n}}{2m(1+p)t}\nonumber \\
&+& mp\frac{k_{i}}{2m(1+p)t}(m-1)p\sum_{n\in
\Omega}\frac{k_{n}}{2m(1+p)t} \nonumber \\
&+& m\frac{k_{i}}{2m(1+p)t}(m-1)p\sum_{n\in
\Omega}\frac{k_{n}}{2m(1+p)t}.
\end{eqnarray}

\begin{figure}
\begin{center}
\includegraphics[width=10cm]{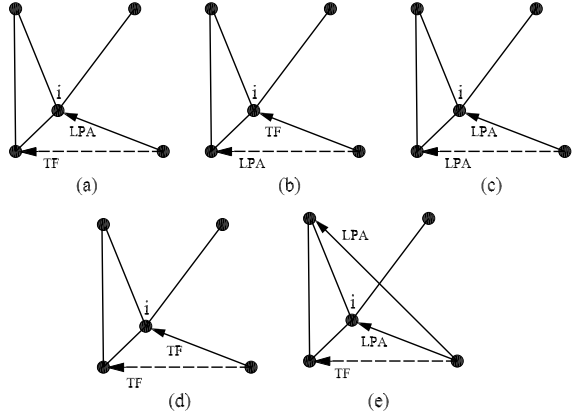}
\caption{The microscopic mechanisms increasing $e_{i}$. The dashed
edges increase $e_{i}$.} \label{f3}
\end{center}
\end{figure}\label{fig:clustering02}

The five terms in the right hand of Eq. (\ref{ei01}) give the
increase in $e_{i}$ by mechanisms from (a) to (e) in turn. It should
be noted that the third term describes mechanism (c) and it is the
only one that would remain if we consider the LC model. In  Eq.
(\ref{ei01}), $k_{i}/[2m(1+p)t]$ is the local preferential
attachment probability to node $i$, $p$ is the triad formation
probability; $k_{n}$ denotes the degree of a neighbor of node $i$,
and $1/k_{n}$ comes from the fact that the neighboring node where a
TF step links is chosen uniformly from the neighbors; $\sum_{n \in
\Omega}k_{n}$ is the sum of the degrees of all neighbors of $i$.

After some simplifications to Eq. (\ref{ei01}), we obtain
\begin{equation}\label{ei02}
\frac{\partial e_{i}}{\partial
t}=\frac{k_{i}}{(1+p)\,t}p+(1+p+p^{2})\frac{k_{i}}{2(1+p)t}(m-1)
\sum_{n\in \Omega} \frac{k_{n}}{2m(1+p)t}.
\end{equation}
In addition, for uncorrelated random networks we have \cite{EgKl02}
\begin{equation}\label{knn}
\sum_{n\in \Omega}{k_{n}}=k_{i}\frac{\langle k\rangle}{4}\ln
t=k_{i}\frac{(1+p)m}{2}\ln t.
\end{equation}
We approximate $e_{i}$ by integrating both sides in Eq.
(\ref{ei02}). The integral for the first term in the right hand of
Eq.~(\ref{ei02}) is simply
\begin{eqnarray}\label{int01}
\int_{1}^{N}
\frac{k_{i}}{(1+p)t}p dt &=&\frac{2p}{1+p}\int_{1}^{N}\frac{dk_{i}}{dt}dt \nonumber\\
&=&\frac{2p}{1+p}\left[k_{i}(N)-m(1+p)\right]\approx \frac{2p}{1+p}
k_{i}(N),
\end{eqnarray}
where we have made use of Eq.~(\ref{ki02}). Using Eqs.~(\ref{ki03})
and~(\ref{knn}), we can integrate the second term in the right hand
of Eq.~(\ref{ei02})
\begin{eqnarray}
&\quad& \quad \int_{1}^{N}(1+p+p^{2})\frac{k_{i}}{2(1+p)t}
(m-1)\sum_{n\in \Omega} \frac{k_{n}}{2m(1+p)t}dt \nonumber\\
&=&(1+p+p^{2})\frac{m-1}{4m(1+p)^{2}}\int_{1}^{N}
\frac{k_{i}^{2}}{t^{2}} \, \frac{m(1+p)}{2}\ln t dt \nonumber\\
&=&(1+p+p^{2})\frac{m^{2}(m-1)(1+p)}{8t_{i}}\left[\frac{(\ln t)^{2}}{2}\right]_{1}^{N} \nonumber\\
&=&(1+p+p^{2})\frac{m-1}{8(1+p)}\frac{(\ln N)^2}{N}k_{i}^{2}(N).
\end{eqnarray}
Combining this with Eq.~(\ref{int01}) yields
\begin{equation}
e_{i}=e_{i,0}+\frac{2p}{1+p}k_{i}(N)+(1+p+p^{2})\frac{m-1}{8(1+p)}\frac{(\ln
N)^{2}}{N}k_{i}^{2}(N).
\end{equation}
Then the clustering coefficient for nodes with large degree $k$
becomes
\begin{equation}\label{ck}
C(k)=\frac{e}{k(k-1)/2}\approx
\frac{4p/(1+p)}{k}+(1+p+p^{2})\frac{m-1}{4(1+p)}\frac{(\ln
N)^{2}}{N},
\end{equation}
after neglecting $n_{i,0}$ (see Fig.~5).
\begin{figure}\label{f4}
\centering
\begin{tabular}{cc}
\subfigure[$m=1,p=0.1$] 
{
    \includegraphics[width=4.2cm]{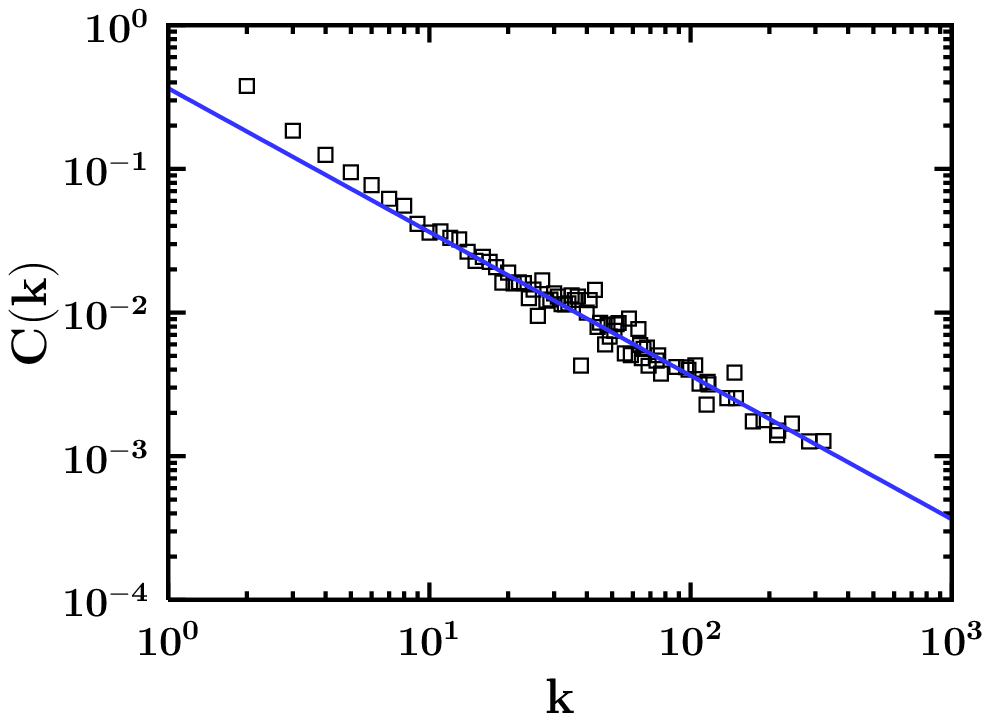}
}
\subfigure[$m=1,p=0.5$] 
{
    \includegraphics[width=4.2cm]{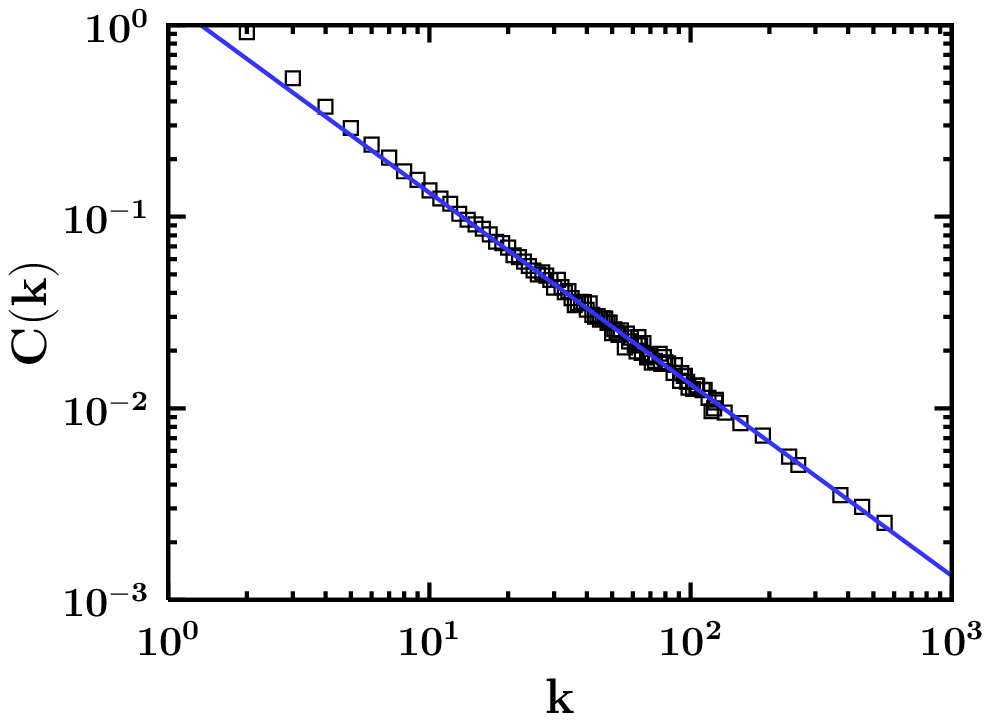}
}
\subfigure[$m=1,p=1$] 
{
    \includegraphics[width=4.2cm]{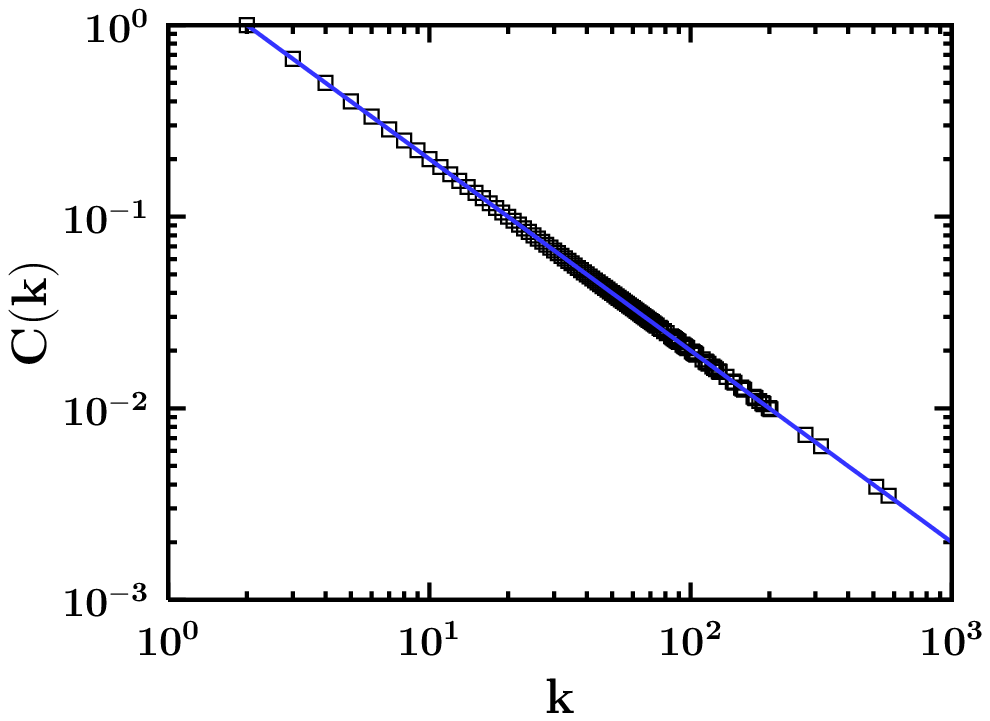}
}\\
\subfigure[$m=3,p=0.1$] 
{
    \includegraphics[width=4.2cm]{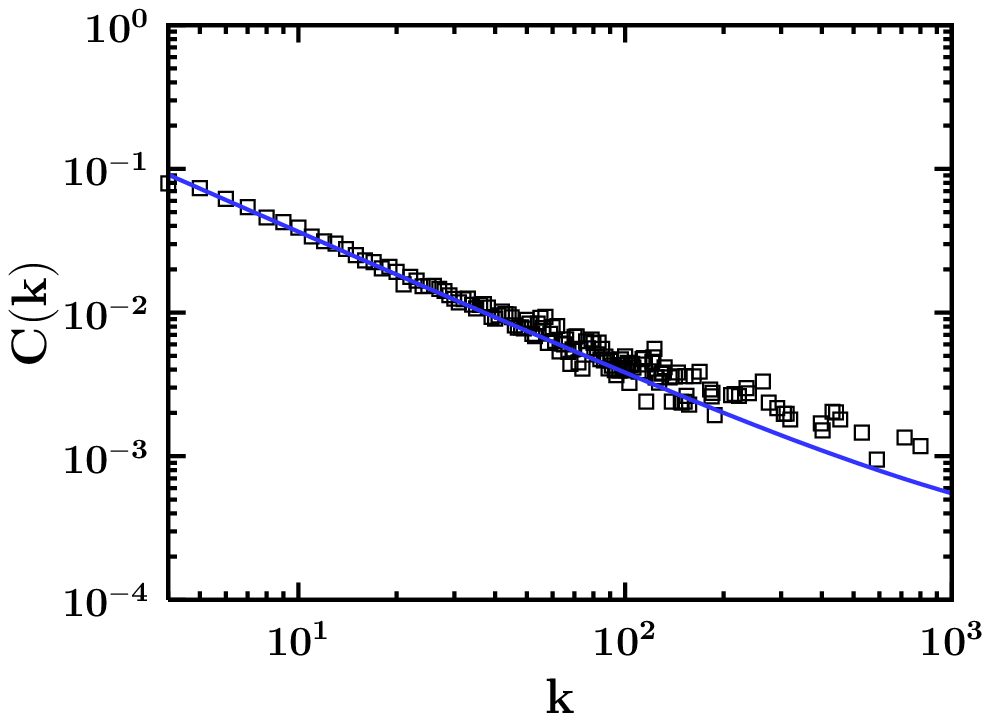}
}
\subfigure[$m=3,p=0.5$] 
{
    \includegraphics[width=4.2cm]{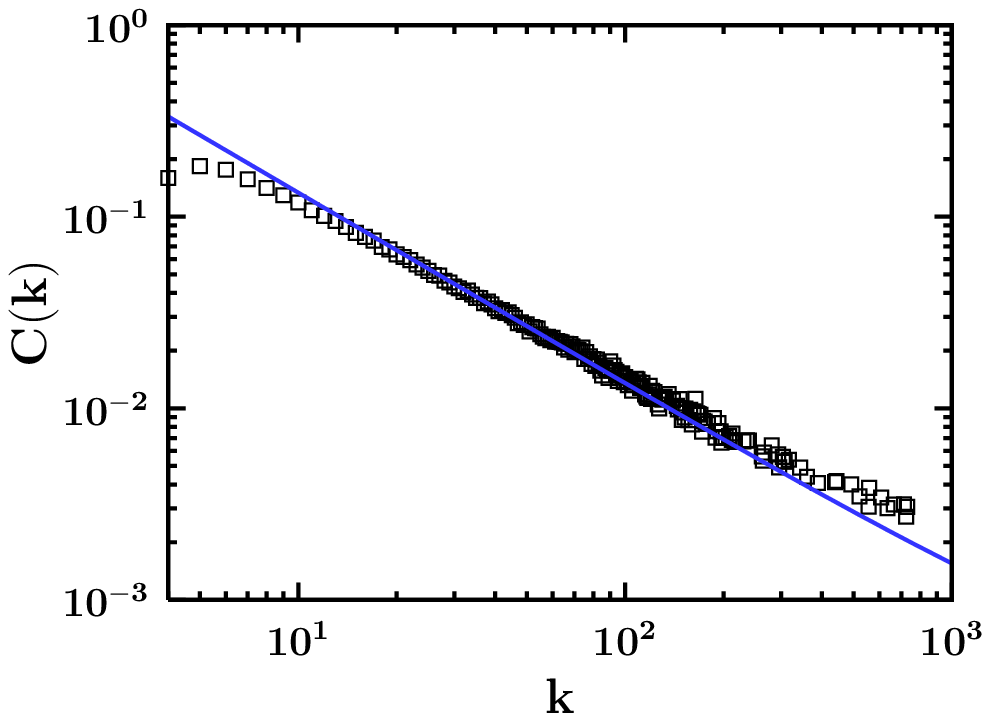}
}
\subfigure[$m=3,p=1$] 
{
    \includegraphics[width=4.2cm]{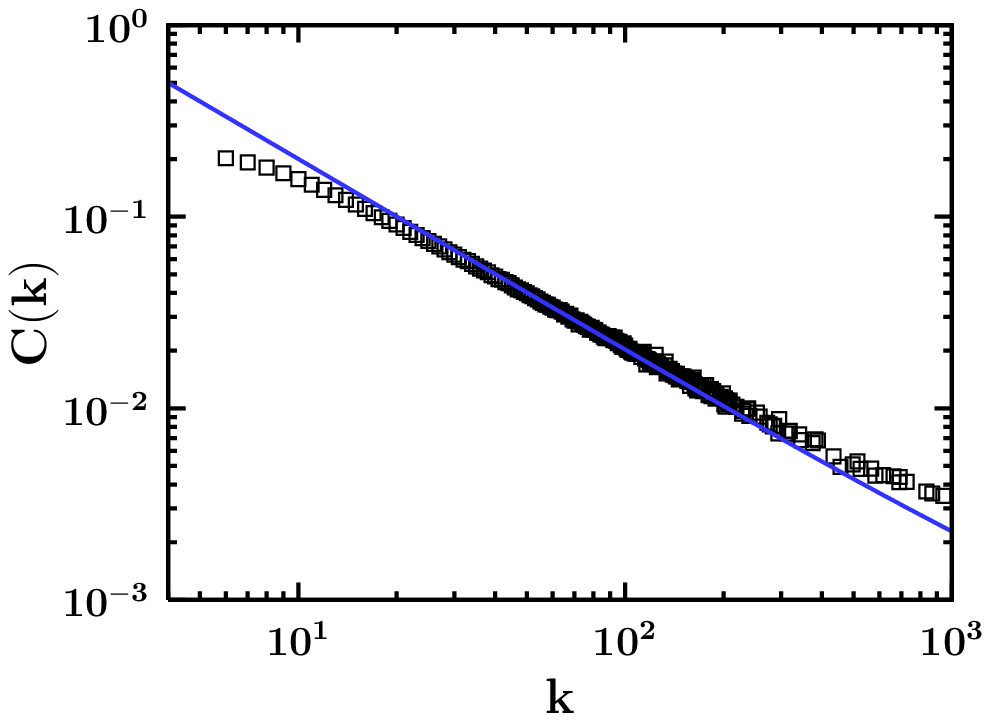}
}\\
\subfigure[$m=5,p=0.1$] 
{
    \includegraphics[width=4.2cm]{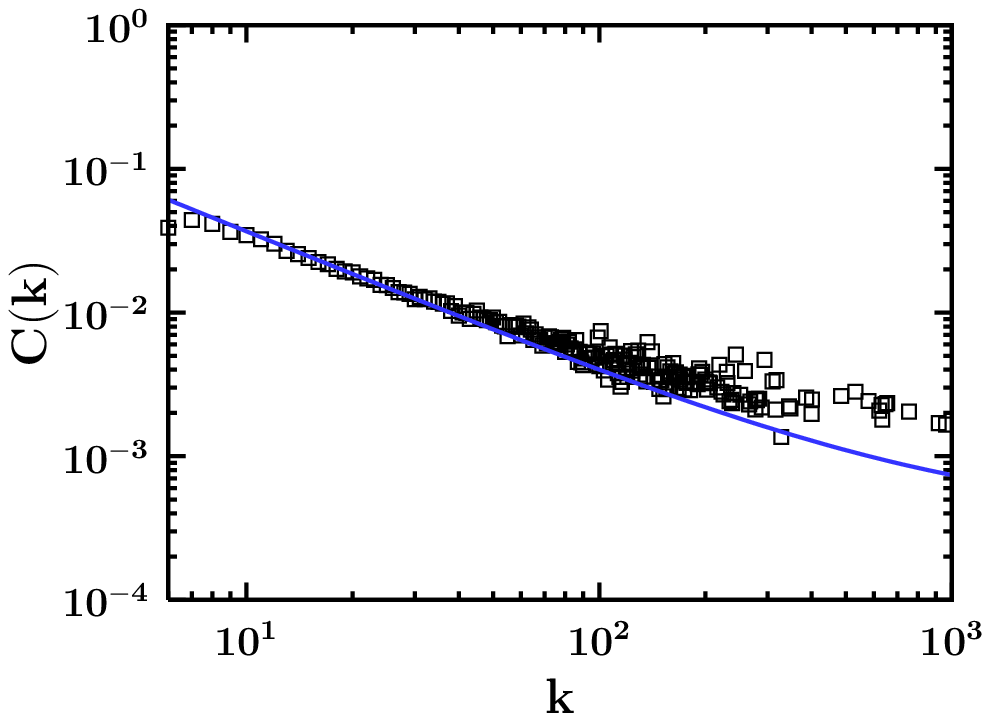}
}
\subfigure[$m=5,p=0.5$] 
{
    \includegraphics[width=4.2cm]{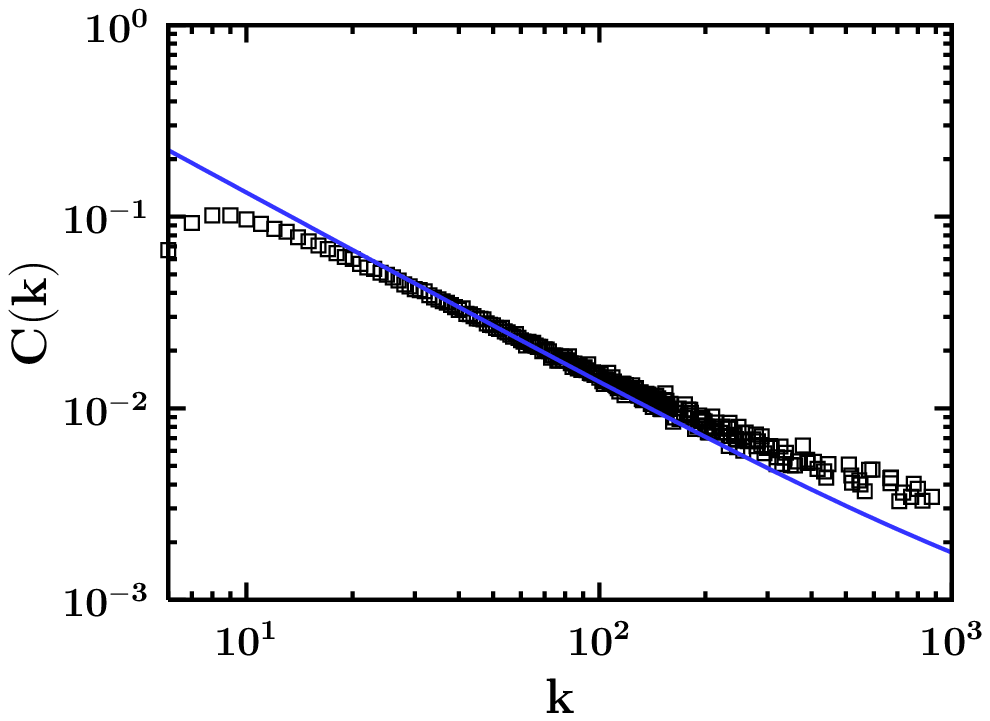}
}
\subfigure[$m=5,p=1$] 
{
    \includegraphics[width=4.2cm]{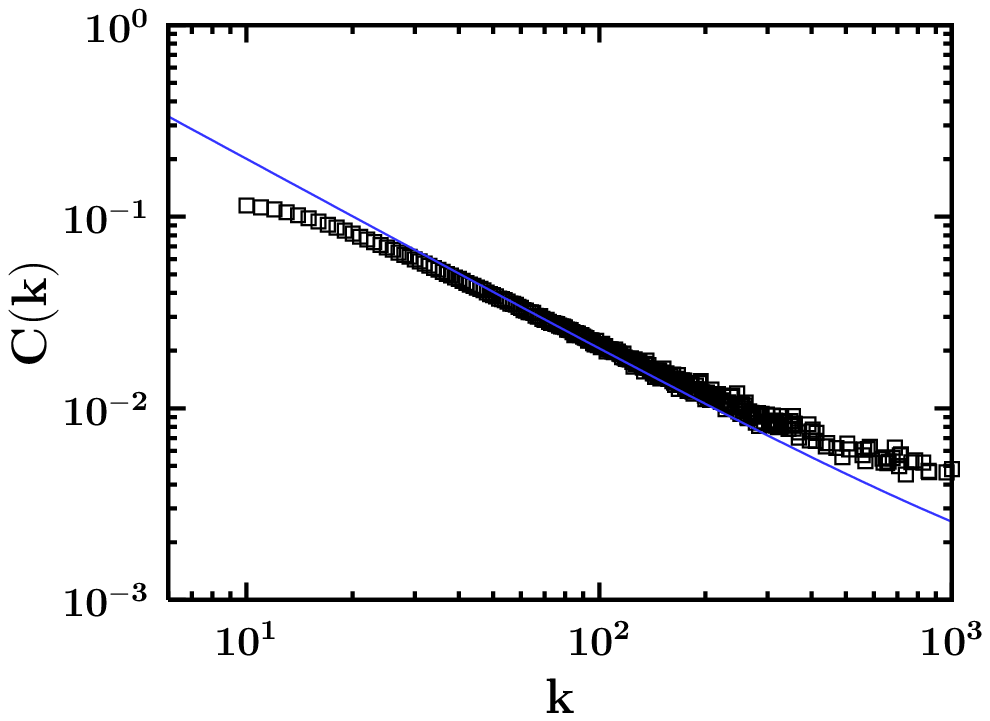}
}
\end{tabular}
\caption{Log-log graphs for clustering coefficient $C(k)$ as a
function of the node degree $k$ for different $m$ and $p$. The sizes
of all networks are 200,000. The local-world size $M_{t}$ scales as
$M_{t}=0.2(m_{0}+t)+m$. The open squares represent the simulation
result and the blue lines are the predictions given by
Eq.~(\ref{ck}). Simulations consistently confirm the analytical
results obtained from the rate equation.}
\end{figure}

Thus, the parameter $p$ in our model introduces the clustering
effect into the system by allowing the formation of triads. By
setting $p$ to a value between 0 and 1 the clustering coefficient of
individual nodes can be adjusted continuously and grows
monotonically with an increasing $p$. In the expression of $C(k)$,
the first term can be ascribed to the triad formation induced
clustering, and shows the $k^{-1}$ behavior that has been observed
in several real-life systems \cite{RaBa03}. This clustering property
is similar to that in other networks such as deterministic (random)
pseudofractal scale-free
networks~\cite{DoGoMe02,CoFeRa04,ZhRoZh07,DoMeSa01,BoPa05} and their
variants~\cite{OzHuOt04,ZhRoGo05,ZhRoCo05a}, the highly clustered
networks on the basis of a finite memory of
nodes~\cite{KlEg02a,KlEg02b}, and Apollonian
networks~\cite{AnHeAnSi05,DoMa05,ZhYaWa05,ZhCoFeRo05,ZhRoCo05,ZhRoZh06}.
Note that $C(k)$ consists of a power law and a constant, so perfect
power-law behavior follows only when the former one dominates.
In the case of $p=0$, we get the clustering coefficient $C(k)$ of
nodes in the LC model
\begin{equation}
C(k)=\frac{m-1}{8} \frac{(\ln N)^{2}}{N},
\end{equation}
which goes to zero as $N$ becomes large enough.

The average clustering coefficient $C$  can be obtained as the mean
value of $C(k)$ with respect to the degree distribution $P(k)$
expressed by Eq.~(\ref{Pk01}). The result is
\begin{eqnarray}\label{ACC01}
 C &=&\sum_{k=m(1+p)}^{\infty} P(k)C(k).
\end{eqnarray}
Although for general $p$ and $m$, it is not easy to derive a closed
formula for the average clustering coefficient $C$, for some
limiting cases we can calculate $C$ analytically. For example, when
both $p$ and $m$ equal 1, equation~(\ref{ACC01}) is reduced to
\begin{eqnarray}\label{ACC02}
 C &=&\sum_{k=2}^{\infty} P(k)C(k)=\sum_{k=2}^{\infty}
 \frac{12}{k(k+1)(k+2)}\,\frac{2}{k} \nonumber \\
&=& \sum_{k=2}^{\infty}
\left(-\frac{18}{k}+\frac{12}{k^{2}}+\frac{24}{k+1}-\frac{6}{k+2}\right)=2\pi^{2}-19\approx0.7392,
\end{eqnarray}
where we have used the fact that $\sum_{m=1}^{\infty}
\frac{1}{m^{2}}=\frac{1}{6\pi^{2}}$. Thus, the average clustering
coefficient is very large.

\subsection{Average path length}
From the above discussions, we find that the existing model shows
both the scale-free nature and the high clustering at the same time.
Moreover, our model exhibits small-word property. Next, we will show
that our network has at most a logarithmic average path length (APL)
with the number of nodes. Here APL means the minimum number of edges
connecting a pair of nodes, averaged over all pairs of nodes.

\begin{figure}
\begin{center}
\includegraphics[width=8cm]{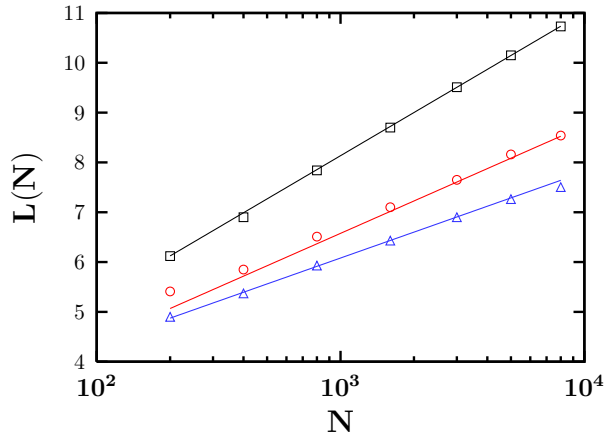}
\caption{Average path length $L(N)$ versus network size $N$ on a
semilogarithmic scale with $m=1$. The open triangles, circles, and
squares denote the cases of $p$=0, 0.5, 1, respectively. The
local-world size $M_{t}$ scales as $M_{t}=0.2(m_{0}+t)+m$. The
straight lines are fits to the data.} \label{f5}
\end{center}
\end{figure}

First, using an approach similar to that presented in
Ref.~\cite{ZhYaWa05}, we study the APL of our network for the
particular case $m=1$ and $p=0$. We label each of the network nodes
according to their creation times, $v=1,2,3,\ldots,N-1,N.$ We denote
$L(N)$ as the APL of our network with size $N$. It follows that
$L(N)=\frac{2\,S(N)}{N(N-1)}$, where $S(N)=\sum_{1 \leq i<j \leq
N}d_{i,j}$ is the total distance, and where $d_{i,j}$ is the
smallest distance between node $i$ and $j$. Note that the distances
between existing node pairs are not affected by the addition of new
nodes. Then we have the following equation:
\begin{equation}\label{APL01}
S(N+1)=S(N)+ \sum_{i=1}^{N}\,d_{i,N+1}.
\end{equation}
As in the analysis of~\cite{ZhYaWa05,ZhRoCo05}, Eq.~(\ref{APL01})
can be rewritten approximately as:
\begin{equation}\label{APL02}
S(N+1) \approx S(N)+N+\frac{2S(N)}{N},
\end{equation}
which implies
\begin{equation}\label{APL03}
\frac{\partial S (N)}{\partial N} =  N + \frac{2S(N)}{N},
\end{equation}
leading to
\begin{equation}\label{APL04}
\varepsilon(N) = N^2(\ln N + \beta),
\end{equation}
where $\beta$ is a constant. When $N$ is large enough, $S(N) \sim
N^2\ln N $, thus we have $L(N) \sim \ln N$. Therefore, we have
presented that in the special case of $m=1$ and $p=0$, there is a
slow growth of the APL with the network size $N$. It should be noted
that in our model, considering values of $m$ greater than 1 and
$p>0$, then the APL will increase more slowly than in the case of
$m=1$ and $p=0$, since in those cases the larger $m$ and $p$ are,
the denser the network becomes. In Fig.~\ref{f5}, we present average
path length $L(N)$ versus network size $N$ with $m=1$ at various
values of $p$. One can see that $L(N)$ increases logarithmically
with $N$.

\section{Conclusions}
In summary, we have presented an expanded local-world evolving
network model with extended power-law degree distribution, a finite
clustering and small average path length. We have obtained the
analytic solutions for relevant network parameters of the considered
model. By changing the expected value $p$ of triad formation steps
after a single LPA, one can tune the clustering coefficient in a
systematic way. In addition, in the evolution of the network, random
fluctuation in the number of new edges is involved which can be
adjusted via tuning $p$. Our model may provide valuable insight into
the real-life networks.

 Although local-world exists
in many real-life networks, it should be pointed out that the choice
of a local-world in real-life world networks is more intricate and
flexible. We use here the most generic case, i.e. random selection,
in our proposed model as in the LC model \cite{LiCh03}. Future work
should include studying in detail the real formation mechanisms of
local-worlds in different real-life networks as well as their
impacts on network topology and dynamics.

\section*{Acknowledgment}
This research was supported by the National Natural Science
Foundation of China under Grant Nos. 60373019, 60573183, and
90612007, and the Postdoctoral Science Foundation of China under
Grant No. 20060400162. L.L. Rong gratefully acknowledges the partial
support from NNSFC under Grant Nos. 70431001 and 70571011. The
authors thank Zhen Shen for his assistance in preparing the
manuscript.


\end{document}